**Abnormal Resistance Switching Behaviors of NiO Thin Films: Possible Occurrence of Simultaneous Formation and Rupture of Conducting Channels**

Chunli Liu, S. C. Chae, S. H. Chang, S. B. Lee, and T. W. Noh[1]

*ReCOE & FPRD, Department of Physics and Astronomy, Seoul National University, Seoul 151-747, Korea*

J. S. Lee, and B. Kahng

*Department of Physics and Astronomy, Seoul National University, Seoul 151-747, Korea*

D.-W. Kim

*Department of Applied Physics, Hanyang University, Ansan, Gyeonggi-Do 426-791, Korea*

C. U. Jung

*Department of Physics, Hankuk University of Foreign Studies, Yongin, Gyeonggi-do 449-791, Korea*

S. Seo and Seung-Eon Ahn

*Samsung Advanced Institute of Technology, Suwon 440-600, Korea*

We report the detailed current-voltage (I-V) characteristics of resistance switching in NiO thin films. In unipolar resistance switching, it is commonly believed that conducting filaments will rupture when NiO changes from a low resistance to a high resistance state. However, we found that this resistance switching can sometimes show abnormal behavior during voltage- and current-driven I-V measurements. We used the random circuit breaker network model to explain how abnormal switching behaviors could occur. We found that this resistance change can occur *via* a series of avalanche processes, where conducting filaments could be formed as well as ruptured.

---

[1] Corresponding author. Electronic mail: twnoh@snu.ac.kr

There has been a great deal of effort to develop new types of nonvolatile memory.[1–3] There have been many reports that resistance values in several binary transition metal oxides[4–7] and perovskite oxides[8–12] could be switched between a high resistance (HR) and a low resistance (LR) state by an external electrical voltage. Resistance switching phenomena have attracted renewed attention due to their potential for creating high-density resistance random access memory (RRAM) devices.[13–18]

Among these resistance switching materials, NiO has been studied extensively because of its high HR/LR ratio and its ease of integration with the available semiconductor process for creating high-density RRAM devices.[13–15] NiO thin films normally show a unipolar resistance switching characteristic, where the resistance can be switched for two values of applied voltage with the same polarity. The unipolar resistance switching in NiO has commonly been explained by the formation and rupture of metallic conducting filaments.[15, 16] Specifically, the SET process, which corresponds to the change from the HR to the LR state, can be explained in terms of the formation of conducting filaments, while the RESET process, which is the change from the LR to the HR state, can be explained by the rupture of conducting filaments.

Recently, we proposed the random circuit breaker (RCB) network model to explain unipolar resistance switching behavior. In this model, we proposed a switching medium composed of circuit breakers that can be switched between bistable resistance states depending on the voltage applied to each component. We demonstrated that the RCB network model could successfully explain most experimental features of unipolar resistance switching.[18]

In this letter, we report abnormal resistance switching behavior during the RESET

processes of NiO capacitors, which cannot be explained simply by the rupture of conducting filaments alone. We observed this anomalous behavior in both voltage- and current-driven measurements. We performed computer simulations based on the RCB network model to explain these intriguing data and discovered that our model can explain the observed abnormal switching behaviors. Our detailed analysis indicated that the RESET process can occur *via* a series of avalanche processes, during which conducting filaments could be formed as well as ruptured.

We fabricated polycrystalline NiO thin films with a thickness of 100 nm on Pt/Ti/SiO$_2$/Si substrates using DC magnetron reactive sputtering. Details of the sample preparation have been reported elsewhere.[13] We measured the electrical properties of the Pt/NiO/Pt capacitors using an HP/Agilent 4155C Semiconductor Parameter Analyzer (Agilent Technologies, Santa Clara, CA). We established a maximum current value for the SET process during voltage-driven measurements. This maximum current limit, which we call the current compliance, was to prevent complete dielectric breakdown during current-voltage (I-V) measurements. For current-driven measurements, we used voltage compliance in the RESET process.

We observed several types of abnormal resistance switching behavior during the RESET process. Figure 1 shows the resistance switching behaviors of our NiO capacitors under DC voltage-driven I-V sweeps. The solid (red) and dashed (black) lines correspond to the RESET and SET processes, respectively. In most cases, resistance switching during the RESET process occurred rather sharply, as shown in Fig. 1(a). However, we sometimes observed a multilevel switching behavior with intermediate states between LR and HR states, as shown in Fig. 1(b). Quite rarely, we observed an even more abnormal resistance switching behavior, where the switching current could become larger than the onset value, denoted by "*s*" in Fig. 1(c). These types of abnormal behavior could not be explained simply by the rupture of conducting filaments occurring during the RESET process. Such abnormal switching behaviors could represent a serious disadvantage for practical applications of RRAM.

We used the RCB network model[18] to explain the abnormal switching behaviors. Figure 2(a) shows a schematic diagram of the RCB network composed of bistable circuit breakers. The resistance value of each circuit breaker will be either $r_l$ for the on-state (*i.e.*, "connected") or $r_h$ for the off-state ("disconnected"). We assume that the state of each circuit breaker can be switched depending on its biased voltage, $\Delta v$, and values of threshold voltages, $v_{\text{off}}$ and $v_{\text{on}}$. As shown schematically in Fig. 2(b), a circuit breaker that is initially in the off-state will be switched to on-state when $\Delta v > v_{\text{on}}$. Conversely, as shown in Fig. 2(c), a circuit breaker that is initially in the on-state will be switched to the off-state when $\Delta v > v_{\text{off}}$. In the LR state, there will be a percolating cluster of on-state circuit breakers; however, such a cluster will not exist in the HR state. Note that $\Delta v$ of each circuit breaker will be determined by the overall applied voltage and status of all the other circuit breakers in the network.

We used two-dimensional square lattices of 150×30 breakers for our simulations, with $r_h/r_l = 1000$ and $v_{\text{on}}/v_{\text{off}} = 9.4$. We randomly set 0.5% of the circuit breakers to the on-state in the pristine network. After increasing the external voltage by a given voltage step, we calculated $\Delta v$ for each circuit breaker and checked the switching conditions shown in Figs. 2(b) and (c). If switching occurred in at least one circuit breaker, we reevaluated the $\Delta v$ distribution and checked the switching conditions again, repeating this iterative process until the network reached a stable state. Then, we increased the external voltage by another voltage step and repeated the calculations until the total current through the network reached its compliance value. The resulting configuration should correspond to the LR state. The details of our simulations have been reported elsewhere.[18]

The percolating network for the LR state and the resulting I-V curve for the RESET process were dependent on the details of the pristine state and the compliance current value. Figures 3(a)–(c) show three snapshots of the RCB network close to a percolating cluster of on-state circuit breakers in such an LR state. Although all the clusters were generated with the same percentage of on-state breakers in the pristine state, the details of the resulting percolating cluster depended very much on the pattern of their random selection. After a percolating cluster was formed, we began to increase the applied voltage from zero again. The I-V curves in Figs. 3(d)–(f) correspond to the initial LR states in Figs. 3(a)–(c). Note that these simulated I-V curves are quite similar to the experimental I-V curves in Figs. 1(a)–(c). This agreement demonstrated that our RCB network model can explain the experimentally observed abnormal switching behaviors quite well.

We examined the regions near the percolating cluster more closely to gain a better understanding of this behavior. As the I-V curve in Fig. 3(f) was the most complicated and most intriguing, we studied the details of the RCB network for the values of *V* marked with the numbers 1–5 in Fig. 3(f). Figure 4 shows detailed snapshots of the portion of the RCB network enclosed by the thick solid (black) box in Fig. 3(c). Note that no configuration changes were observed in other parts of the network during this simulation. Solid (black) and dashed (blue) lines are used in Fig. 4 to indicate regions where rupture and formation of conducting filaments occur, respectively, to clarify the progressive changes in the RCB network. Although this simulation result corresponds to a RESET process, we found that rupture and formation of conducting filaments could occur simultaneously. The switching of at least one circuit breaker could bring about a significant change in $\Delta v$ distribution, which would lead to an avalanche of switchings in other nearby circuit breakers. The close agreement between Figs. 1(c) and 3(f) indicates that rupture and formation of conducting filaments may not be exclusive processes.

We also observed similar abnormal RESET switching behavior in current-driven switching measurements. Note that in current-driven I-V sweeps, the voltage across the capacitor is determined by the values of the applied current and the capacitor resistance. Figure 5(a) shows a typical I-V curve in current-driven measurements with a voltage compliance ($V_{COMP}$) of 1.0 V. With a small current, the network is in the LR state, and so the slope of the I-V curve is large. After the RESET process, the network is in the HR state, and so its slope has a much smaller value. When $V_{COMP}$ was increased to 2.0 V, we observed the abnormal RESET switching shown in Fig. 5(b), which exhibits several resistance value changes. Note that, after the first change, the resistance became lower than that of the original LR state while it was still undergoing the RESET process. When $V_{COMP}$ was increased further to 2.5 V, we observed a series of abnormal switchings before the capacitor experienced a hard-breakdown, as shown in Fig. 5(c). Kim *et al.* reported a similar series of abnormal switchings in current-driven measurements for the SET process.[11]

We found that the RCB network model can also explain these abnormal switching behaviors in the current-driven I-V measurements. Figures 5(d)–(f) show I-V curve simulations for current-driven resistance switching with increasing values of $V_{COMP}$. Note that these simulation results agreed quite closely with the experimental results. We also studied the details of snapshots for the abnormal switchings, and found that rupture and formation of conducting filaments could also occur simultaneously during current-driven measurements.

In summary, we observed abnormal resistance switching behavior during the RESET process in NiO thin film capacitors, a phenomenon that can pose a serious obstacle for practical applications of RRAM. We found that the RCB network model can explain how such abnormal resistance switching can occur. The success of the RCB model suggests that it will be possible to use this percolation-type of model to find ways of suppressing abnormal resistance switching.

This work was supported financially by the Creative Research Initiatives (Functionally Integrated Oxide Heterostructure) of the Korean Science and Engineering Foundation (KOSEF).

**Figure captions**

Fig. 1. (color online) I-V characteristics of NiO capacitors during SET and RESET processes for DC voltage-driven I-V sweeps. The dashed (black) and solid (red) lines represent the SET and RESET processes, respectively. (a) Typical resistance switching. (b) Abnormal multilevel resistance switching with an intermediate state. (c) Abnormal resistance switching with large resistance fluctuations during the RESET process. "*s*" indicates the onset of the RESET process where the resistance value begins to change.

Fig. 2. (color online) (a) Schematic diagram of the random circuit breaker network model for a two-dimensional bond percolation. Each bond represents one circuit breaker. The red and green colors represent on- and off-states, respectively. The upper and lower black bars represent top and bottom electrodes. (b) Change in the circuit breaker's state from off to on. (c) Change in the circuit breaker's state from on to off.

Fig. 3 (color online) Simulation results for voltage-driven RESET processes using the RCB network model. All the simulations were performed in 150× 30 square lattices of circuit breakers with 0.5% of the circuit breakers randomly turned on in the pristine state. (a), (b), (c) Snapshots of the RCB network near a percolating cluster of on-state circuit breakers in the LR state. Note that the thick (red) and thin (green) lines correspond to on- and off-state circuit breakers, respectively. (d), (e), (f) Voltage-driven I-V sweep results for the configurations shown in (a), (b), and (c), respectively.

Fig. 4. (color online) Snapshots of circuit breaker network configuration at each voltage, labeled in the I-V curve in Fig. 3(f). In each figure, the regions where conducting

filaments are formed and ruptured are marked by the dashed (blue) and solid (black) lines, respectively.

Fig. 5. (a), (b), (c) Experimental I-V curves for RESET processes under current-driven I-V sweeps with $V_{COMP}$ = 1.0, 2.0, and 2.5 V, respectively. (d), (e), (f) Simulated I-V curves with different compliance voltages values of 2, 6, and 10, respectively. Note that $V_{COMP}$ values in the simulation are unitless.

**Figure 1**

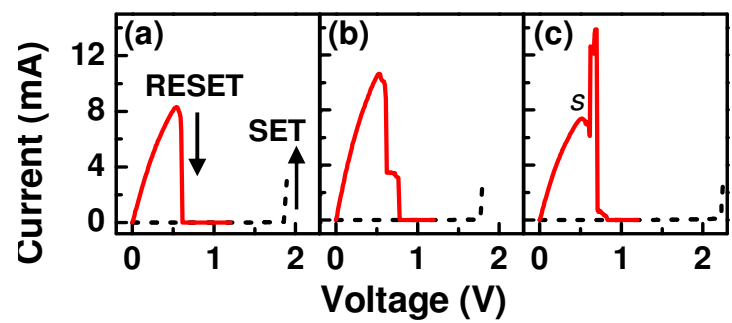

**Figure 2**

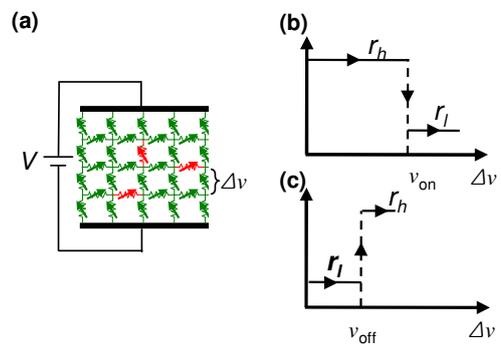

**Figure 3**

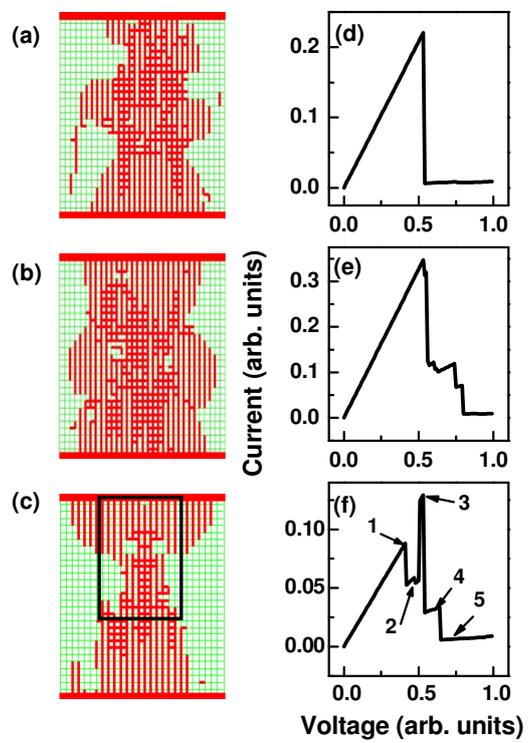

**Figure 4**

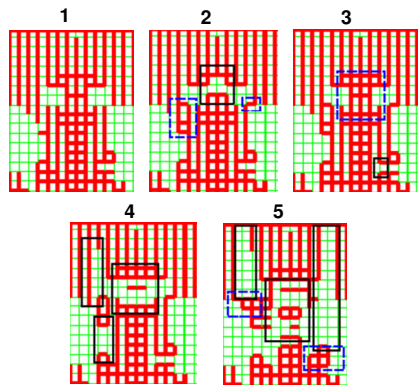

**Figure 5**

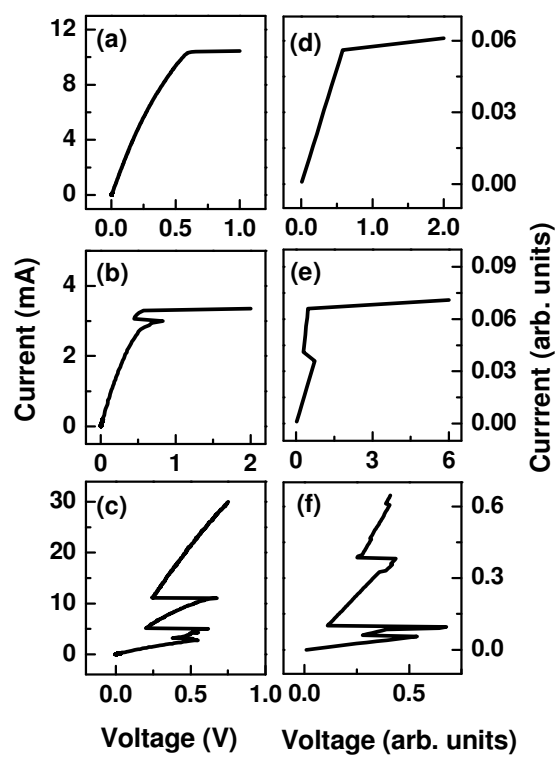